\documentclass[12pt]{iopart}
\usepackage{epsf}
\begin{document}

\title[Systematical Approach to the Exact Solution of the Dirac Equation...]{Systematical Approach to the Exact Solution of the Dirac
Equation for A Special Form of the Woods-Saxon Potential}

\author{C\"uneyt Berkdemir\dag\footnote[1]{(berkdemir@erciyes.edu.tr)}, Ay\c se Berkdemir\dag\footnote[3]{(arsland@erciyes.edu.tr)}
and Ramazan Sever\ddag\footnote[2]{(sever@metu.edu.tr)}}

\address{\dag\ Department of Physics, Faculty of Arts and
Sciences, Erciyes University, Kayseri, 38039, Turkey}

\address{\ddag\ Department of Physics, Middle East Technical
University, Ankara, 06531, Turkey}

\begin{abstract}

Exact solution of the Dirac equation for a special form of the
Woods-Saxon potential is obtained for the s-states. The energy
eigenvalues and two-component spinor wave functions are derived by
using a systematical method which is called as Nikiforov-Uvarov.
It is seen that the energy eigenvalues strongly depend on the
potential parameters. In addition, it is also shown that the
non-relativistic limit can be reached easily and directly.

\end{abstract}


\submitto{\JPA}

\maketitle

\section{Introduction}
The Woods-Saxon potential and its various modifications have been
received for much interest describing metallic clusters in a
successful way. It is used in the central part of the interaction
neutron with one heavy-ion nucleus and also for the optical
potential model \cite {ref1}. Differential cross-section in the
$^{16}O+ ^{12}C$ elastic scattering are analyzed in some energies
by using this potential \cite {ref2}. The quantum behavior of the
relativistic particle in the presence of a central Woods-Saxon
potential has settled the possible existence of bound state
spectra. The relativistic Dirac-oscillator and
Dirac-exponential-type potential problems have already been
established by adding an off-diagonal linear radial term to the
Dirac operator for a long time ago \cite {ref5,ref6}. Recently,
the relativistic bound states spectrum and its eigenfunctions for
the triaxial and axially deformed harmonic oscillators have been
derived as well \cite {ref7}. In addition, a mixture of the Dirac
oscillator (tensor potential) with vector and scalar harmonic
oscillator potentials has been solved analytically for the general
case \cite {ref8}.

Furthermore, only a few articles for the relativistic problems
have been written on the Dirac equation with the exponential-type
potential. The Dirac equation has been solved by making use of
two-component spinors for the exponential type potentials such as
Woods-Saxon and Hulth\'{e}n potentials for a special case. Kennedy
has studied the generalized approach to the Woods-Saxon potential
and obtained the scattering and bound-state solutions of the
one-dimensional Dirac equation. However, more realistic cases have
not been discussed \cite {ref9}. A. D. Alhaidari has just
introduced a new formalism to the definition of the radial Dirac
equation and solved for a class of shape-invariant potentials
\cite {ref10,ref11,ref12}. The main point in the formalism is that
two coupled first order differential equations resulting from the
radial Dirac equation generate Schr\"odinger-like equations for
the two spinor components. Following the procedure given in
Ref.\cite {ref10}, we present a new systematical approach to solve
the Dirac-Woods-Saxon problem by means of the Nikiforov-Uvarov
(NU) method \cite {ref13}. The non-relativistic limit reproduces
the well-known non-relativistic energy spectrum and results in the
Schr\"odinger equation for a special form of the Woods-Saxon
potential \cite {ref14}.

The article is structured as follows: In Section 2, we briefly
introduce an overview of the technical details of the formalism
improved by Alhaidari. After that, the basic concepts of the
Nikiforov-Uvarov method are given in the same section to solve the
Dirac-Woods-Saxon problem. Section 3 is devoted to the solution of
the problem to obtain the energy eigenvalues and eigenfunctions by
applying the NU method. The paper is concluded with a short
summary in Section 4.

\section{Formalism and Method}

We shall first introduce the Alhaidari's formalism proposed to
solve the Dirac equation for spherically symmetric potential
interactions. Later, the fundamental mathematical relations of the
NU method will be summarized to obtain the solution of the
Schr\"odinger-like equations easily and systematically.

\subsection{Overview of the Formalism}

Alhaidari's formalism is based on the consideration by writing the
relativistic Hamiltonian for a Dirac spinor coupled to a
four-component electromagnetic potential $(A_0, \vec{A})$. After
using Gauge invariance and spherically symmetric of the
electrostatic potential, free Dirac equation transforms to the
matrix representation of the Dirac Hamiltonian (see Ref.\cite
{ref10} for more detail). For convenience, atomic units are
selected as $m=e=\hbar=1$ and the speed of light $c$ is matched
with $\alpha^{-1}$. Thus, the Hamiltonian for a Dirac spinor in
four-component electromagnetic potential $(A_0, \vec{A})$ can be
written as follows:
\begin{equation}
\label{eq1} H =  \left( \begin{array}{cc} 1+\alpha A_0  & -i\alpha
\vec{\sigma}.\vec{\nabla}+i\alpha \vec{\sigma}.\vec{A}
\\\\
-i\alpha \vec{\sigma}.\vec{\nabla}-i\alpha \vec{\sigma}.\vec{A}
 & -1+\alpha A_0
\end{array} \right )
\end{equation}
where $\alpha$ is the fine structure parameter and $\vec{\sigma}$
are the three 2x2 Pauli spin matrices. Taking the spherically
symmetric case and writing $(A_0, \vec{A})$ as $(\alpha V(r),
\hat{r}W(r))$, the two-component Dirac equation is obtained as
\begin{equation}
\label{eq2} H =  \left(\begin{array}{cc} 1+\alpha^2V(r)-E_R  &
\alpha \left[\frac{\kappa}{r}+W(r)-\frac{d}{dr}\right]
\\\\
\alpha \left[\frac{\kappa}{r}+W(r)+\frac{d}{dr}\right]
 & -1+\alpha^2V(r)-E_R
\end{array} \right)\left(\begin{array}{cc} g(r)\\\\ f(r)
\end{array} \right)=0
\end{equation}
where $f(r)$ and $g(r)$ are real radial square integrable functions,
$E_R$ is the relativistic energy and $\kappa$ is the spin-orbit
coupling parameter defined as $\kappa=\pm (j+1/2)=\pm 1, \pm 2, ...$
for $l=j\pm 1/2$. However, the current problem is analytically
solvable only for $\ell =0$ (s-states). In addition, $V(r)$ and
$W(r)$ are the even and odd components of the relativistic
potential, respectively. For a given value of the spin-orbit
coupling parameter $\kappa$, Schr\"odinger-like requirement relates
the two potential functions as:
$W(r)=\frac{1}{\xi}V(r)-\frac{\kappa}{r}$, where $\xi$ is a real
parameter and $V(r)$ is not depend on the $\kappa$ parameter. In
order to obtain the Schr\"odinger-like equation in the formalism
proposed by A. D. Alhaidari, it is used a global unitary
transformation which eliminates the first derivative. Thus
$\mathcal{U}(\eta)=exp(\frac{i}{2}\alpha \eta \sigma_2)$ is applied
in Eq.(\ref{eq2}). Here, $\eta$ is a real constant and $\sigma_2$ is
the 2x2 Pauli matrix which defines the two radial spinor components
in terms of the other,
\begin{equation}
\label{eq3} \phi^\mp(r)= \frac{\alpha}{C\pm E_R}\left[-\xi
\pm\frac{C}{\xi}V(r)+\frac{d}{dr}\right]\phi^\pm(r),
\end{equation}
with $C=cos(\alpha \eta)=\sqrt{1-(\alpha \xi)^2}>0$,
\begin{equation}
\label{eq4}
\left(\begin{array}{cc} \phi^+(r)\\
\phi^-(r)
\end{array} \right)=\mathcal{U}\left(\begin{array}{cc} g(r)\\
f(r)
\end{array} \right).
\end{equation}
Here $\phi^{\pm}(r)$ is the upper or lower spinor components
respectively. It is emphasized that Eq.(\ref{eq3}) with the top
and bottom signs are not valid for negative and positive energy
solutions respectively. The top and bottom signs in front of $E_R$
in Eq.(\ref{eq3}) are not allowed to take the values $-C$ and
$+C$. Because these two values are elements of the negative and
positive energy spectra respectively. Substituting these into the
radial Dirac equation (Eq.(\ref{eq2})), we get the
Schr\"odinger-like second-order differential equation in terms of
the lower and upper spinor components as
\begin{equation}
\label{eq5} \left[-\frac{d^2}{dr^2}+\frac{C^2}{\xi^2}V^2+2E_RV \mp
\frac{C}{\xi}\frac{dV}{dr}-\frac{E_R^2-1}{\alpha^2}\right]\phi
^\pm(r)=0,
\end{equation}
where the "+" sign belongs to the upper spinor component, while
the other sign corresponds to the lower one.

\subsection{Basic Concepts of the Method}

Solutions of the Schr\"odinger-like second order differential
equations play an essential role in studying many important
problems of theoretical physics. In this point, the NU method can
be used to solve these types equations with an appropriate
coordinate transformation $s=s(r)$ \cite {ref13}:
\begin{equation}
\label{eq6} \psi ^{\prime \prime }(s)+\frac{\stackrel{\sim }{\tau
}(s)}{\sigma (s) }\psi ^{\prime }(s)+\frac{\stackrel{\sim }{\sigma
}(s)}{\sigma ^{2}(s)}\psi (s)=0
\end{equation}
where $\sigma (s)$ and $\stackrel{\sim }{\sigma }(s)$ are
polynomials with at most second-degree, and $\stackrel{\sim }{\tau
}(s)$ is a first-degree polynomial. It is of fundamental
importance in the study of particular special orthogonal
polynomials \cite {ref15}. These polynomials try to reduced
Eq.(\ref{eq6}) to a simple form by taking $\psi(s)=\phi(s)y(s)$
and choosing an appropriate $\phi(s)$. Consequently,
Eq.(\ref{eq6}) can be reduced to an equation of hypergeometric
type
\begin{equation}
\label{eq9} \sigma(s)y^{\prime \prime }(s)+\tau(s)y^{\prime
}(s)+\lambda y(s)=0,
\end{equation}
where $\tau(s)=~\stackrel{\sim }{\tau}(s)+2\pi (s)$ (its
derivative must be negative) and $\lambda$ is a constant, which is
given in the form
\begin{equation}
\label{eq8} \lambda =\lambda _{n}=-n\tau ^{\prime
}-\frac{n(n-1)}{2}\sigma ^{\prime \prime },~~~(n=0,1,2,...).
\end{equation}
Here, $\lambda$ or $\lambda_n$ are obtained from a particular
solution of the form $y(s)=y_n(s)$ which is a polynomial of degree
$n$. $y_n(s)$ is the hypergeometric type function whose polynomial
solutions are given by Rodrigues relation
\begin{equation}
\label{eq10}
y_{n}(s)=\frac{B_{n}}{\rho (s)}\frac{d^{n}}{ds^{n}}\left[ \sigma ^{n}(s)\rho (s)%
\right] ,
\end{equation}
where $B_{n}$ is the normalization constant and the weight
function $\rho(s)$ must be satisfied the condition
\begin{equation}
\label{eq11} [\sigma(s)\rho(s)]^{\prime }=\tau(s)\rho(s).
\end{equation}
To determine the weight function given in Eq.(\ref{eq11}), we must
immediately obtain the polynomial $\pi(s)$ from:
\begin{equation}
\label{eq13} \pi =\frac{\sigma ^{\prime }-\stackrel{\sim }{\tau
}}{2}\pm
\sqrt{\left(\frac{\sigma^{\prime}-\stackrel{\sim}{\tau}}{2}\right)^{2}-\stackrel{\sim}{\sigma}+k{\sigma}}.
\end{equation}
In principle, the expression under the square root sign in
Eq.(\ref{eq13}) can be arranged as the square of a polynomial.
This is possible only if its discriminant is zero. In this case,
it is obtained an equation for $k$. After solving this equation,
the obtained values of $k$ are included to the NU method and here
there is a relationship with $\lambda$ of $k$ so that
$k=\lambda-\pi^{\prime}(s)$. After this point, an appropriate
$\phi(s)$ can be invented from $\phi (s)^{\prime }/\phi (s)=\pi
(s)/\sigma (s)$.

\section{Special form of the Woods-Saxon potential}
The interaction between nuclei is commonly described by using a
potential that consists of the Coulomb and the nuclear potentials.
It is usually taken in the form of the Woods-Saxon potential.
Here, we take the following special form for the Woods-Saxon
potential which is specified by "$q$" parameter
\begin{equation}
\label{eq15}
V(r)=-\frac{qV_{0}}{q+e^{\left(\frac{r-R_{0}}{b}\right)}}~,
\end{equation}
where $V_0$ is the potential depth, $R_{0}$ is the width of the
potential, $b$ is thickness of the surface which is usually
adjusted to the experimental values of ionization energies and $q$
is a real positive parameter which is responsible for the
specification of the Woods-Saxon potential. After substituting the
potential into Eq.(\ref{eq5}), we obtain an equation for the upper
spinor component
\begin{eqnarray}
\label{eq16} \fl
\left[-\frac{d^2}{dr^2}+\frac{C^2}{\xi^2}\left(\frac{qV_{0}}{q+e^{\left(\frac{r-R_{0}}{b}\right)}}\right)^2-\frac{2qE_RV_{0}}{q+e^{\left(\frac{r-R_{0}}{b}\right)}}
- \frac{qCV_0}{\xi
b}\frac{e^{\left(\frac{r-R_{0}}{b}\right)}}{\left(q+e^{\left(\frac{r-R_{0}}{b}\right)}\right)^2}-\frac{E_R^2-1}{\alpha^2}\right]\phi
^+(r)=0,\nonumber\\
\left[-\frac{d^2}{dr^2}+\frac{qCV_0}{\xi
b}\frac{\frac{qCV_0b}{\xi}-e^{\left(\frac{r-R_{0}}{b}\right)}}{\left(q+e^{\left(\frac{r-R_{0}}{b}\right)}\right)^2}-\frac{2qE_RV_{0}}{q+e^{\left(\frac{r-R_{0}}{b}\right)}}
-\frac{E_R^2-1}{\alpha^2}\right]\phi ^+(r)=0.~~~~~~~~~~
\end{eqnarray}
In order to apply the NU$-$method, we rewrite Eq.(\ref{eq16}) by
using a new variable of the form
$s=-e^{-\left(\frac{r-R_{0}}{b}\right)}$,
\begin{equation}
\label{eq17} \fl
\left[-\frac{s}{b}\frac{d}{ds}\left(\frac{s}{b}\frac{d}{ds}\right)+\frac{q^2C^2V_0^2}{\xi^2}\left(\frac{s}{1-qs}\right)^2+\frac{2qsE_{R}V_{0}}{1-qs}
+ \frac{qCV_0}{\xi
b}\frac{s}{\left(1-qs\right)^2}-\frac{E_R^2-1}{\alpha^2}\right]\phi
^+(s)=0.
\end{equation}
By introducing the following dimensionless parameters
\begin{equation}
\label{eq18}
\varepsilon=\left(\frac{E_R^2-1}{\alpha^2}\right)b^2,~~~~~~\beta=2qE_RV_0b^2,~~~~~~\gamma=\frac{qCV_0b}{\xi},
\end{equation}
we reach the following hypergeometric type equation defined in
Eq.(\ref{eq6})
\begin{equation}
\label{eq19} \fl
\frac{d^2\phi^+(s)}{ds^2}+\frac{1-qs}{s(1-qs)}\frac{d\phi^+(s)}{ds}~
+\frac{1}{s^2(1-qs)^2}\times\left[(q^2\varepsilon+q\beta-\gamma^2)s^2-(2q\varepsilon+\beta
+\gamma)s+\varepsilon\right]\phi^+(s)=0.
\end{equation}
After comparing Eq.(\ref{eq19}) with Eq.(\ref{eq6}), we obtain the
corresponding polynomials:
\begin{equation}
\label{eq20} \stackrel{\sim}{\tau }(s)=1-qs,~~~{\sigma
}(s)=s(1-qs),~~~\stackrel{\sim}{\sigma
}(s)=(q^2\varepsilon+q\beta-\gamma^2)s^2-(2q\varepsilon+\beta
+\gamma)s+\varepsilon.
\end{equation}
Substituting these polynomials into Eq.(\ref{eq13}), we organize
the polynomial $\pi(s)$ as follows
\begin{equation}
\label{eq21} \fl \pi (s)=-\frac{qs}{2}\pm
\frac{1}{2}\sqrt{\left(q^2-4(q^2\varepsilon+q\beta-\gamma^2)
-4qk\right)s^2+4\left(2\varepsilon q+\beta +\gamma
+k\right)s-4\varepsilon},
\end{equation}
with $ \sigma ^{\prime }(s)=1-2qs$. It is taken into consideration
that the discriminant of the second order equation under the
square root sign has to be zero. Hence, the expected roots are
obtained as $k_{\pm}=\left(-\beta -\gamma \pm i (q+2\gamma
)\sqrt{\varepsilon}\right)$. In this case, substituting these
values for each \textit{k} into Eq.(\ref{eq21}), the possible
solutions are obtained for $\pi (s)$
\begin{equation}
\label{eq22} \pi(s) = -\frac{qs}{2}\pm
\frac{1}{2}\left\{\begin{array}{cc}
\left(q+2\gamma-2iq\sqrt{\varepsilon}\right)s+2i\sqrt{\varepsilon},\\
\hskip 0.1cm \mbox{for} \hskip 0.2 cm k_+=-\beta -\gamma +i
(q+2\gamma)\sqrt{\varepsilon}\\\\
\left(q+2\gamma+2iq\sqrt{\varepsilon}\right)s-2i\sqrt{\varepsilon},\\
\hskip 0.1cm \mbox{for} \hskip 0.2 cm k_-=-\beta -\gamma -i
(q+2\gamma)\sqrt{\varepsilon}\\
\end{array}\right.
\end{equation}
From the four possible forms of the polynomial $\pi (s)$, we take
a certain one which is the derivative of $\tau(s)$ has a negative
value. Therefore, the function $\tau (s)$ satisfies the following
equalities:
\begin{eqnarray}
\tau(s)=1+2i\sqrt{\varepsilon}-s\left(3q+2\gamma+2iq\sqrt{\varepsilon}~\right),\nonumber
\end{eqnarray}
\begin{equation}
\label{eq23}
\tau^{\prime}(s)=-\left(3q+2\gamma+2iq\sqrt{\varepsilon}~\right).
\end{equation}
In the present case
\begin{equation}
\label{eq24} \pi(s)=-\frac{qs}{2}-\frac{1}{2}\left[\left(q+2\gamma
+2iq\sqrt{\varepsilon}~\right)s-2iq\sqrt{\varepsilon}~\right].
\end{equation}
From $k= \lambda -\pi^{\prime}(s)$ and also Eq.(\ref{eq8}), we
obtain respectively:
\begin{equation}
\label{eq25}
\lambda=-\beta-2\gamma-q-2i(\gamma+q)\sqrt{\varepsilon}
\end{equation}
\begin{equation}
\label{eq26}
\lambda=\lambda_n=n^2q+2nq+2n\gamma+2niq\sqrt{\varepsilon}.
\end{equation}
After having the comparison of Eq.(\ref{eq25}) and Eq.(\ref{eq26})
and substituting the values of $\varepsilon$ and $\beta$, we can
immediately obtain the $\kappa$-independent relativistic energy
eigenvalues $E_{Rnq}$ of the Dirac particle as follows
\begin{equation}
\begin{array}{l}
\label{eq27}
 \fl E_{Rnq}^ \pm = -[2b(T^2+b^2q^2V_0^2\alpha^2)]^{ - 1} \{ (1 + n) q V_0 b (T+\gamma) \alpha ^2  \pm [(1 + n)^2(T+\gamma)^2 q^2V_0b^2
 \\\\
\hskip 3.5cm -
(T^2+b^2q^2V_0^2\alpha^2)((1+n)^2(T+\gamma)^2-4b^2T^2)]^{{1
\mathord{\left/ {\vphantom {1 2}} \right.
\kern-\nulldelimiterspace} 2}} \},  \\
\end{array}
\end{equation}
where $T=(1+n)q+\gamma$. To have a physical result, the expression
under the square root must be positive. $n$ is a positive integer
defined in the interval of $n_{max}\geq n\geq 0$ and is called the
radial quantum number.

By interesting with Eq.(\ref{eq16}), we can easily show that, in
the non-relativistic limit $\alpha \rightarrow 0$, the
relativistic energy is a limit of the non-relativistic energy,
$E_R\approx 1+\alpha^2 E_{NR}$, where $E_{NR}$ is the
non-relativistic energy. The wave equation is reduced to the
following form, choosing $q=1$:
\begin{eqnarray}
\label{eq28}
\left[-\frac{d^2}{dr^2}+\frac{\gamma}{b^2}\frac{\gamma-e^{\left(\frac{r-R_{0}}{b}\right)}}
{\left(1+e^{\left(\frac{r-R_{0}}{b}\right)}\right)^2}-\frac{2V_{0}}{1+e^{\left(\frac{r-R_{0}}{b}\right)}}
-2E_{NR}\right]\phi ^+(r)=0.
\end{eqnarray}
To obtain a more suitable case, we can use the following form
after taking $\gamma= -1$,
\begin{eqnarray}
\label{eq29}
\left[-\frac{d^2}{dr^2}-2\frac{V_{0}-1/2b^2}{1+e^{\left(\frac{r-R_{0}}{b}\right)}}
-2E_{NR}\right]\phi ^+(r)=0,
\end{eqnarray}
which is in the form of the Schr\"odinger equation for the new
type s-wave non-relativistic Woods-Saxon potential. The
corresponding energy spectrum has already been given in Ref. \cite
{ref3} by using the hypergeometric functions, and also repeated by
means of the NU method in Ref. \cite {ref4} as:
\begin{equation}
\label{eq30}
E_{NR}=-\frac{1}{2}\left[\frac{b(V_0-1/2b^2)}{n+1}+\frac{n+1}{2b}\right]^2,
\end{equation}
where the transformation $V_0\rightarrow V_0-1/2b^2$ is applied for
the convenience and the index $n$ relates to the radial quantum
number ($n=0, 1, 2, 3,...$). In order to obtain the relativistic
energy spectrum directly, considering the relativistic
Eq.(\ref{eq16}) and the non-relativistic Eq.(\ref{eq30}) for the
case of $\gamma =-1$, we can propose the relevant parameter map:
\begin{eqnarray}
\label{eq31} b\rightarrow b,~~~~~~~~R_0\rightarrow R_0,\nonumber\\
V_0-1/2b^2\rightarrow E_{R}V_0-1/2b^2,\nonumber\\
E_{NR}\rightarrow (E_{R}^2-1)/2\alpha^2.
\end{eqnarray}
Using the map between the parameters of the two equations, the
resulting upper relativistic energy spectrum is found as follows:
\begin{equation}
\label{eq32} \fl E_{R}^+=\frac{-V_0b\alpha^2((n+1)^2-1)
+\sqrt{4b^2(n+1)^4+(n+1)^2\alpha^2[4b^4V_0^2-((n+1)^2-1)^2]}}{2b((n+1)^2+b^2V_0^2\alpha^2)}.
\end{equation}
It can be easily seen that the relativistic energy spectrum in
Eq.(\ref{eq27}) gives the same result as Eq.(\ref{eq32}) but it
can be used only for the s-states. In addition, Eq.(\ref{eq27})
indicates that one deals with a family of the Woods-Saxon
potential and the relativistic energy spectrum will be also used
to describe the single-particle motion in nuclei. Therefore,
Eq.(\ref{eq27}) can also give the solution of the relativistic
Dirac-Woods-Saxon problem with a general value of $\gamma$ for
$q=1$.

Let us now find the corresponding wave functions. According to the
NU-method, the polynomial solutions of the hypergeometric function
$y(s)$ depend on the determination of weight function $\rho(s)$
satisfying the differential equation $[\sigma (s)\rho (s)
]^{\prime }=\tau (s)\rho (s)$. Thus, $\rho(s)$ is calculated as
\begin{equation}
\label{eq33}
\rho(s)=\left(1-qs\right)^{\nu}s^{2i\sqrt{\varepsilon}},
\end{equation}
where $\nu=1+\frac{2\gamma }{q}$. Substituting into the Rodrigues
relation given in Eq.(\ref{eq10}), the wave functions are obtained
in the following form
\begin{equation}
\label{eq34}
y_{nq}(s)=A_{n}\left(1-qs\right)^{-\nu}s^{-2i\sqrt{\varepsilon}}\frac{d^{n}}{ds^{n}}\left[\left(1-qs\right)^{n+\nu}s^{n+2i\sqrt{\varepsilon}}%
\right],
\end{equation}
where $A_{n}$ is the normalization constant. Taking $q=1$, the
polynomial solutions of $y_n(s)$ are expressed in terms of the
Jacobi Polynomials, which is one of the orthogonal polynomials. In
this case, the weight function is
$(1-s)^{\nu}s^{2i\sqrt{\varepsilon}}$ and Eq.(\ref{eq34}) is
reduced to $\sim P_n^{(2i\sqrt{\varepsilon},~\nu)}(1-2s)$ \cite
{ref15}. After substituting $\pi(s)$ and $\sigma(s)$ into the
expression $\phi (s)^{\prime }/\phi (s)=\pi (s)/\sigma (s)$, the
other part of the wave function is found as
\begin{equation}
\label{eq35} \phi(s)=(1-qs)^{\nu-\gamma
/q}s^{i\sqrt{\varepsilon}}.
\end{equation}
We write the upper spinor component in terms of the Jacobi
polynomials
\begin{equation}
\label{eq36}
\phi_n^+(s)=B_ns^{i\sqrt{\varepsilon}}(1-s)^{\nu-\gamma}P_n^{(2i\sqrt{\varepsilon},~\nu)}(1-2s),
\end{equation}
where $B_n$ is a normalization constant. The lower component of
the spinor wave function can also be obtained by substituting
Eq.(\ref{eq36}) into Eq.(\ref{eq3}). We should then solve the
following equation
\begin{equation}
\label{eq37} \phi_n^-(s)= \frac{\alpha}{C+ E_{Rn}^\pm}\left[-\xi
+\frac{C}{\xi}V(s)-\frac{s}{b}\frac{d}{ds}\right]\phi_n^+(s).
\end{equation}
where $E_{Rn}^\pm\neq C$. This is possible if a new variable is
introduced as $x=1-2s$ . Now, the equation of the lower spinor
component has been transformed into the following form
\begin{equation}
\label{eq38} \phi_n^-(s)= \frac{\alpha}{C+ E_{Rn}^\pm}\left[-\xi
+\frac{V_0C(1-x)}{\xi(1+x)}+\frac{(1-x)}{b}\frac{d}{dx}\right]\phi_n^+(s),
\end{equation}
with
\begin{equation}
\label{eq39}
\phi_n^+(s)=C_n(1-x)^{i\sqrt{\varepsilon}}(1+x)^{\nu-\gamma}P_n^{(2i\sqrt{\varepsilon},~\nu)}(x),
\end{equation}
where $C_n$ is the normalization constant and its value is equal
to $B_n2^{\gamma-\nu-i\sqrt{\varepsilon}}$. If the following
recursion relations and the differential formula satisfied by the
Jacobi polynomials \cite {ref16} are included to the solution
\begin{eqnarray}
\fl (1+x)P_n^{(\mu,~
\varrho)}(x)=\frac{2}{2n+\mu+\varrho+1}\left[(n+\varrho)P_n^{(\mu,~
\varrho-1)}(x)+(n+1)P_{n+1}^{(\mu,~\varrho-1)}(x)\right],\nonumber
\end{eqnarray}
\begin{eqnarray}
\fl (1-x)P_n^{(\mu,~
\varrho)}(x)=\frac{2}{2n+\mu+\varrho+1}\left[(n+\mu)P_n^{(\mu-1,~
\varrho)}(x)-(n+1)P_{n+1}^{(\mu-1,~\varrho)}(x)\right],\nonumber
\end{eqnarray}
\begin{eqnarray}
\fl (1-x^2)\frac{dP_n^{(\mu,~
\varrho)}}{dx}(x)=-n\left(x+\frac{\varrho-\mu}{2n+\mu+\varrho}\right)P_n^{(\mu,~
\varrho)}(x)+2\frac{(n+\mu)(n+\varrho)}{2n+\mu+\varrho}P_{n-1}^{(\mu,~\varrho)}(x),\nonumber
\end{eqnarray}
\begin{eqnarray}
\fl P_n^{(\mu,~
\varrho)}(x)=\frac{n+\mu+\varrho+1}{2n+\mu+\varrho+1}P_n^{(\mu,~
\varrho+1)}(x)+\frac{n+\mu}{2n+\mu+\varrho+1}P_{n-1}^{(\mu,~\varrho+1)}(x),
\end{eqnarray}
we obtained the lower spinor component in terms of the Jacobi
polynomials as a function of $s(r)$
\begin{equation}
\begin{array}{l}
\label{41} \phi_n^-(s)=
\frac{B_n}{2n+\mu+\varrho+1}\frac{\alpha}{C+
E_{Rn}^\pm}(1-s)^\gamma s^{i\sqrt{\varepsilon}}
\{(L+n/b)[(n+\mu)P_{n}^{(2i\sqrt{\varepsilon}-1,~\nu)}(1-2s)
\\\\
\hskip 1.0cm \fl
-(n+1)P_{n+1}^{(2i\sqrt{\varepsilon}-1,~\nu)}(1-2s)]+2(n+\varrho)(2n+\mu+\varrho+1-M)P_{n}^{(2i\sqrt{\varepsilon},~\nu-1)}(1-2s)
\\\\
\hskip 3.5cm
-(1/2b+(n+2b)(2n+\mu+\varrho+1)/2)P_{n}^{(2i\sqrt{\varepsilon},~\nu)}(1-2s)
\},
\end{array}
\end{equation}
where $L=CV_0/\xi+(1+\gamma)/b$ and
$M=\xi+i\sqrt{\varepsilon}/b$.\\

\section{Conclusion}
We have solved the Dirac equation for the special form of the
Woods-Saxon potential following a formalism introduced by
Alhaidari. The NU method is used to obtain a systematical solution
in the Dirac-Woods-axon problem. The energy spectrum of the bound
states is analytically obtained and two-component spinor
eigenfunctions are written in terms of the Jacobi polynomials. It
is seen that the energy eigenvalues are a function of the
parameter q and the solution space splits into two distinct
subspaces. We have seen that the non-relativistic limit of the
Dirac equation can be obtained easily. We can also say that the
exact results obtained for a special form of the Woods-Saxon
potential give us some interesting applications in
the various quantum mechanical studies and the relativistic nuclear scattering problems.\\

\noindent {\bf Acknowledgements}

The authors are indebted to A. D. Alhaidari for the useful
comments on the original version of the manuscript. This research
was partially supported by the Scientific and Technological
Research Council of Turkey.

\end{document}